\documentclass{article}
\usepackage{epsfig,bbm,graphicx}
\usepackage{graphics,amsmath,amsfonts,revsymb,latexsym}
\newtheorem{theorem}{Theorem}[section]
\newtheorem{corollary}[theorem]{Corollary}
\newtheorem{lemma}[theorem]{Lemma}

\newtheorem{proposition}[theorem]{Proposition}

\newcommand{\qed}{{\hfill$\Box$}}
\newenvironment{proof}{\noindent \textbf{{Proof~} }}{\qed}
\newenvironment{remark}{\noindent \textbf{{Remark~} }}

\def\bi{\begin{itemize}}
\def\ei{\end{itemize}}
\def\be{\begin{equation}}
\def\ee{\end{equation}}
\def\bea{\begin{eqnarray}}
\def\eea{\end{eqnarray}}
\def\ben{\begin{eqnarray*}}
\def\een{\end{eqnarray*}}

\def\>{\rangle}
\def\<{\langle}

\def\ra{\rightarrow}

\def\eps{\epsilon}

\def\bbE{\mathbb{E}}

\def\bbR{\mathbb{R}}

\newcommand{\eq}[1]{Eq.~(\ref{eq:#1})}

\newcommand{\markov}{{\,\rightarrow \,}}

\newcommand{\bra}[1]{\langle #1 |}
\newcommand{\ket}[1]{| #1 \rangle}

\newcommand{\proj}[1]{| #1 \>\!\< #1 |}

\DeclareMathOperator{\id}{id}

\DeclareMathOperator{\tr}{Tr}

\def\*{\star}

\def\tilde{\widetilde}
\def\bar{\overline}

\def\cA{{\cal A}}		 \def\cB{{\cal B}}		 

\def\cD{{\cal D}}		 \def\cE{{\cal E}}		 

		 \def\cH{{\cal H}}

		 \def\cT{{\cal T}}		 

		 \def\cW{{\cal W}}		 
\def\cX{{\cal X}}
\def\cY{{\cal Y}}

\begin{document}

\title{{Channel simulation with quantum side information}
\protect\vspace{5mm}}
\author{Zhicheng Luo\protect\\
\it{Department of Physics,} 
\it{University of Southern California,}\protect\\
\it{Los Angeles, CA 90089, USA}
\protect\\
\\
Igor Devetak%\footnote{\tt devetak@usc.edu}
\protect\\
\it{Department of Electrical Engineering--Systems,} 
\it{University of Southern California,}\protect\\
\it{Los Angeles, CA 90089, USA}} 
%\date{\today}
\maketitle

\begin{abstract}
We study and solve the problem of classical channel simulation with quantum side information
at the receiver. This
is a generalization of both the classical reverse Shannon theorem, 
and the classical-quantum Slepian-Wolf problem. The optimal noiseless 
communication rate is found to be
reduced from the mutual information between the channel input and output
by the Holevo information between the channel output and the quantum side information. 

Our main theorem has two important corollaries.
The first is a quantum generalization of the Wyner-Ziv problem: rate-distortion theory 
with quantum side information. The second is an alternative proof of the 
trade-off between classical communication and common randomness distilled from a quantum
state. 
%Simple proof of the both direct coding theorems can be made based on our result. The formula for the trade-off
%between the one-way communication invested and the distilled common randomness also follows from our theorem.

The fully quantum generalization of the problem considered is \emph{quantum state 
redistribution}.
Here the sender and receiver share a mixed quantum state and the sender wants to transfer part 
of her state to the receiver using entanglement and quantum communication.
We present outer and inner bounds on the achievable rate pairs.
%a conjecture on the achievable rate pairs which, if true,
%would generalize both the fully quantum Slepian-Wolf (FQSW) and fully quantum 
%reverse Shannon (FQRS) theorems.

\end{abstract}

\section{Introduction}

In his seminal 1948 paper~\cite{Shannon48} Shannon introduced the problem of data compression.
He found that a memoryless source consisting of a large number $n$ of symbols 
generated according to a probability distribution $p$ 
can be compressed without loss at a rate of $H(p)$ bits per symbol, 
where $H(p)$ is the Shannon entropy of $p$. 
This result can be rephrased as a communication problem.
The sender Alice wants to communicate her source to the receiver Bob. 
Equivalently, she wants to {simulate} a noiseless bit channel (which we denote by $\bar{\id}$) 
from her to Bob with respect to the input $p$. She can accomplish this task by using
up a rate $H(p)$ of perfect bit channels (which we denote by $[c \rightarrow c]$) 
from her to Bob. The protocol consists of Alice sending the compressed
source and Bob performing decompression upon receipt. 
The existence of such a protocol may be succinctly expressed as 
a \emph{resource inequality}~\cite{DHW05,DW03a,DHW03}
$$
H(p) \, [c \rightarrow c] \geq \< \bar{\id} : p \>.
$$
The non-local resource on the left hand side can be composed with local
pre- and post-processing to simulate the non-local resource on the right
hand side. 

With this viewpoint in mind, Shannon's result was generalized some 50 years later
to simulating noisy channels. The latter result was dubbed the reverse Shannon
theorem ~\cite{BSST01,Winter02a}, referring to Shannon's noisy \emph{channel coding} theorem 
\cite{Shannon48}.
One may well ask why one should be interested in simulating noise.
The reason is a saving in resources: part of the classical communication
$[c \rightarrow c]$ can be replaced by shared coins or
``common randomness'' (denoted by $[c \, c]$). Common randomness is a strictly
weaker resource than classical communication because Alice can flip her coin locally
and send the outcome to Bob. The reverse Shannon theorem is intimately related \cite{Winter02a}
 to lossy compression, or rate-distortion theory \cite{Berger71}, 
where the communication rate is traded off against a suitably
defined distortion level of the data. More generally, the reverse Shannon theorem is a useful tool for
effecting trade-offs between resources \cite{HJW02, BHLSW03} .

Another generalization of Shannon's result, introduced by Slepian and Wolf \cite{SW73},
is to give Bob side information about source. The case of quantum side information
was considered in \cite{DW02}.

In this paper we combine the two ideas of making the channel noisy and 
allowing quantum side information with the receiver. We also
analyze several consequences for trade-offs. The first is 
rate-distortion theory with quantum side information paralleling
the classical work of Wyner and Ziv~\cite{WZ76}. 
The second is an alternative derivation of a result from ~\cite{DW03a} concerning
distillation of common randomness from a bipartite quantum state with the assistance of
one-way classical communication. 
%We give an alternative
%derivation of the trade-off between classical communication 
%and common randomness distilled. 
The various implications of
our result are shown in Figure 1.

This paper is organized as follows. 
In Section 2 we introduce the notation and give some background.
Section 3 contains our main result, Theorem \ref{thm1}, together with its proof. 
%Our result is rephrased as a resource inequality \cite{DHW05}.
Section 4 discusses consequences of Theorem \ref{thm1}.
In section 5 we find outer and inner bounds for a fully quantum version of our problem.
Section 6 concludes with a discussion and proposed future work.
%Some useful notation and results are collected in the Appendix.

\begin{figure}%[htbp]
\centerline{ {\scalebox{0.7}{\includegraphics{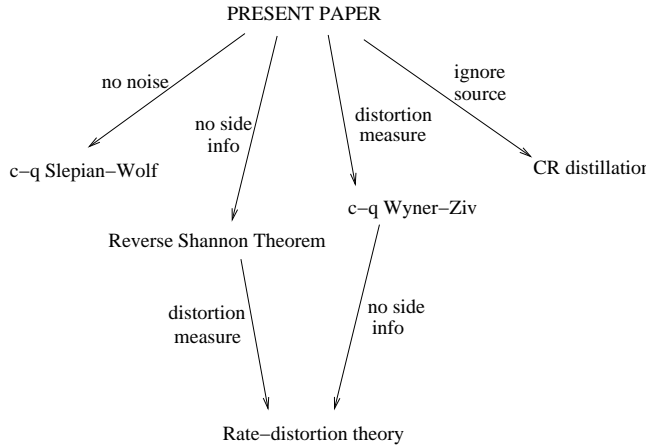}}}}
  \caption{The relation of our results to prior work.}
\end{figure}

\section{Notation}
Let us  introduce some useful notation for the bipartite classical-quantum systems.
The state of a classical-quantum system $XB$ can be described by an ensemble
$\cE = \{\rho^{B}_x, p(x)\}$, with $p(x)$ defined on $\cX$ and the $\rho^{B}_x$ being density
operators on the Hilbert space $\cH_B$ of $B$. Thus, with probability $p(x)$ the classical index
and quantum state take on values $x$ and $\rho^{B}_x$, respectively.
A useful representation of classical-quantum systems is obtained by 
embedding the random variable 
$X$ in some quantum system, also labelled by $X$. 
Then our ensemble $\{\rho^{B}_x, p(x)\}$ corresponds to 
the density operator
\begin{equation}
 \rho^{XB}= \sum_x p(x)\ket{x}\bra{x}^{X}\otimes\rho^{B}_x \label{cq},
\end{equation}  
where $\{\ket{x}: x\in\cX\}$ is an orthonormal basis for the Hilbert space $\cH_{X}$ of $X$.
A classical-quantum system may, therefore, be viewed as a special case of a quantum one. The 
von Neumann entropy of a quantum system $A$ with density operator $\sigma^{A}$ is defined as
$H(A)_\sigma= - \tr{\sigma^{A}\log{\sigma^{A}}}$. The subscript is often omitted.
 For a tripartite quantum system $ABC$ 
in some state $\sigma^{ABC}$ define 
the conditional von Neumann entropy
$$
H(B|A)= H(AB)-H(A) ,
$$ 
quantum mutual information
$$ 
I(A;B)= H(A) + H(B) - H(AB)= H(B)- H(B|A), 
$$ 
and quantum conditional mutual information
$$ 
I(A;B|C)= I(A;BC) - I(A;C).
$$ 
For classical-quantum correlations (\ref{cq}) the von Neumann entropy $H(X)_\rho$ 
is just the Shannon entropy 
$H(X)= -\sum_{x}p(x)\log{p(x)}$ of the random variable $X$. 
The conditional entropy $H(B|X)$  equals $\sum_xp(x)H(\rho^{B}_x)$. 
The mutual information $I(X;B)$ is the Holevo quantity~\cite{Holevo73} of 
the ensemble $\cE$:
$$
\chi(\cE)= H\left(\sum_x p(x)\rho_x\right)- \sum_x p(x)H(\rho_x).
$$ 
Finally we need to introduce a classical-quantum analogue of a \emph{Markov chain}.
We may define a classical-quantum
Markov chain $Y\!\markov\!X\markov\!B$ associated with an ensemble
$\{ \rho^B_{xy}, p (x,y) \}$ for which $\rho^B_{xy} = \rho^B_x$ is independent of $y$.
Such an object 
typically comes about by augmenting the system $XB$ by the
random variable $Y$ (classically) correlated with $X$ via a conditional
distribution $W(y|x) = \Pr\{Y=y|X=x\}$. 
This corresponds to the state
\begin{equation}
\rho^{XYB} = \sum_x p(x) \sum_y W(y|x) \ket{y}\bra{y}^Y 
 \otimes \ket{x}\bra{x}^X \otimes \rho_x^B .\label{ccq}
\end{equation}
Here $W(y|x)$ is the noisy channel and $X$ and $Y$ are input and output
random variables. Therefore the classical-quantum system $YB$ can be expressed as
\begin{equation}
\rho^{YB} = \sum_y q(y) \ket{y}\bra{y}^Y \otimes \rho_y^B
\end{equation} 
with $q(y)= \sum_x p(x)W(y|x)$ and $\rho_y^B = \sum_x P(x|y)\rho_x^B$.

\section{Channel simulation with quantum side information}

Consider a classical-quantum system $XB$ in the state (\ref{cq}) 
such that the sender Alice possesses the classical index $X$ and the receiver Bob has the
quantum system $B$. Consider a classical channel
from Alice to Bob given by the conditional probability distribution 
$W$. Applying this channel to the $X$ part of $\rho^{XB}$ results in the
state $\rho^{XYB}$ given by (\ref{ccq}). 
Ideally, we are interested in  \emph{simulating} the channel $W$ using 
noiseless communication and common randomness,
in the sense that the simulation produces the state $\rho^{XYB}$.
For reasons we will discuss later, we want Alice to also get 
a copy $\bar{Y}$ of the output, so that the final state produced is
\begin{equation}
\rho^{XY\bar{Y}B} = \sum_x p(x) \sum_y W(y|x) \ket{y}\bra{y}^Y 
 \otimes \ket{y}\bra{y}^{\bar{Y}} \otimes  
 \ket{x}\bra{x}^X \otimes \rho_x^B .\label{cccq}
\end{equation}
The systems $X$ and $\bar{Y}$ are in Alice's possession, while
Bob has $B$ and $Y$.

As usual in information theory, this task is amenable to analysis when we go to the 
approximate, asymptotic i.i.d. (independent, identically distributed) setting. 
This means that Alice and Bob share $n$ copies of the classical-quantum system $XB$,
given by the state 
\be
\label{eq:stayte}
\rho^{X^nB^n} = \sum_{x^n} p^n(x^n) \proj{x^n}^{X^n} \otimes \rho^{B^n}_{x^n},
\ee
where $x^n = x_1 \dots x_n$ is a sequence in $\cX^n$,
$p^n(x^n) = p(x_1) \dots p(x_n)$,
 and
$\rho_{x^n}= \rho_{x_1}\otimes\rho_{x_2}\cdots\otimes\rho_{x_n}$.
They want to simulate the channel $W^n(y^n|x^n) = 
W(y_1|x_1) \dots W(y_n|x_n)$
approximately, with error approaching
zero as $n \rightarrow \infty$.
They have access to a rate of $C$ bits/copy  of common randomness, which means that
they  have the same string $l$ picked uniformly at random from the set 
$\{0,1\}^{n C}$. In addition, they are allowed a rate of $R$ bits/copy of classical communication,
so that Alice may send an arbitrary string $m$ from the set $\{0,1\}^{n R}$ to Bob.

An $(n,R,C,\eps)$ simulation code consists of 
\begin{itemize}  
  \item An encoding stochastic map $E_n : \cX^n \times \{0,1\}^{n C} \ra \{0,1\}^{n R} \times \{0,1\}^{n S}$.
If the value of the common randomness is $l \in  \{0,1\}^{n C}$, Alice encodes her classical message 
    $x^n$ as the index $ms$, $m \in\{0,1\}^{n R}$, $s \in\{0,1\}^{n S}$, with probability 
    $E_l(m,s|x^n):=  E_n(m,s|x^n,l)$, and only sends $m$ to Bob;

   \item  A set $\{\Lambda^{(lm)}\}_{lm\in\{0,1\}^{n (C+R)}}$, where each 
    $\Lambda^{(lm)}=\{\Lambda^{(lm)}_{s'}\}_{s'\in\{0,1\}^{n S}}$ is a POVM acting on $B^n$ and 
	  taking on values $s'$. Bob does not get sent the true value of $s$ and needs to
infer it from the POVM;

  \item A deterministic decoding map $D_n : \{0,1\}^{n C}\times\{0,1\}^{n R}\times\{0,1\}^{n S} \ra\cY^n$;
this allows Alice and Bob to produce their respective  simulated outputs $\tilde{y}^n = D_l(m,s):= D_n(l,m,s)$
and $\hat{y}^n = D_l(m,s')$, based on $l$, $m$ and $s$ (in Bob's case $s'$); 
\end{itemize}
such that
\be
\|(\rho^{XY\bar{Y}B})^{\otimes n} - \sigma^{X^n\hat{Y}^n\tilde{Y}^n\hat{B}^n} \|_1 \leq \eps.
\label{41/2}
\ee
Here the state $\sigma^{X^n\hat{Y}^n\tilde{Y}^n\hat{B}^n}$ denotes the 
result of the simulation, which includes Alice's original $X^n$, the post-measurement system $\hat{B}^n$,
Alice's simulation output random variable $\tilde{Y}^n$ and Bob's simulation output random variable $\hat{Y}^n$
(based on $s'$).

%The \emph{rate} $R$, noiseless communication rate, signifies the number of bits per copy needed to encode 
%the index $m$. And \emph{rate} $C$, common randomness consumption rate, indicates the number of bits per 
%simulation needed to determine the index of the simulation code.
A rate pair $(R,C)$ is called achievable if for all $\epsilon>0$, $\delta>0$ and sufficiently large $n$,
there exists an $(n, R+ \delta, C + \delta, \epsilon)$ code.

We now state our main theorem.

\begin{theorem}
\label{thm1}
The region of achievable $(R,C)$ pairs is given by
$$  
R \geq I(X;Y)-I(Y;B), \quad \quad C + R \geq H(Y|B).
$$
\end{theorem}

The theorem contains a direct coding part (achievability) and a converse part (optimality).
For the direct coding theorem it suffices to prove the achievability of the rate pair 
$(R,C) = (I(X;Y)-I(Y;B), H(Y|X))$. The full region given by Theorem \ref{thm1} (see Figure 2) 
follows by observing that a bit of common randomness may be generated from a bit of communication.

\begin{figure}%[htbp]
\label{fig1}
\centerline{ {\scalebox{0.7}{\includegraphics{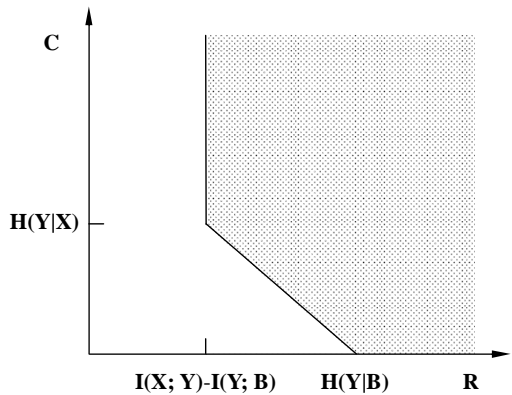}}}}
  \caption{Achievable region of rate pairs for a classical-quantum system $XB$.}
\end{figure}

A naive simulation would be for Alice to actually perform the channel $W$
locally and send a compressed instance of the output to Bob. This would require a communication
rate of $H(Y)$ bits per copy. The first idea is to split this information 
into an intrinsic and extrinsic part \cite{Winter01a}. The extrinsic part has 
rate $H(Y|X)$ and is provided by the common randomness. Only the intrinsic part
$I(X;Y) = H(Y) - H(Y|X)$ requires classical communication. This protocol
would amount to sending the strings $m$ and $s$ above. However, a further savings of
$I(Y;B)$ is accomplished by Bob deducing the $s$ index from his quantum state.
Thus Alice need only send $m$ which requires a rate $I(X;Y) - I(Y;B)$.

For the direct coding part we will need several lemmas.
The first one is the Chernoff bound (cf. \cite{AW02}).

\begin{lemma}[Chernoff bound]
Let $Z_1, \dots, Z_n$ be i.i.d. random variables with mean $\mu$.
Define $\bar{Z}_n = \frac{1}{n}\sum_{j =1}^n Z_j$. If the $Z_j$ take values in the interval $[0 , b]$, 
then for $\eta \leq \frac{1}{2}$, and some constant $\kappa_0$,
\begin{equation}
\Pr\{ |\bar{Z}_n - \mu | \geq \mu \eta \} \leq 2 \exp (- \kappa_0 n \mu \eta^2/b).
\end{equation}
\label{eq:chernoff}
\end{lemma}

The second lemma concerns deterministically ``diluting'' a uniformly distributed random variable to
a non-uniform one on a larger set. We will need it to create $y^n$ from
 $l$, $m$ and $s$.

\begin{lemma}[Randomness dilution]
\label{lemma:rd}
We are given a probability distribution $q(y)$ defined on  $\cY$ and a set $\cT \subseteq \cY$
 such that
\begin{align}
q(\cT)  &:= \sum_{y\in\cT}q(y) \geq 1 - \epsilon,  \label{eq:dil1}\\
q(y) & \geq \alpha, \,\,\, \forall y \in \cT, \label{eq:dil2}
\end{align} 
for some positive numbers $\alpha$ and $\epsilon$.
Let $W$ be the random variable uniformly distributed
on $\{1,\ldots,M\}$.
%\begin{enumerate}
%\item
For random variables $Y_1,Y_2,...,Y_M$ all distributed
according to $q$, define the map $G: \{1,...,M\} \rightarrow \cY$  by $G(i) = Y_i$. 
Then, letting $\tilde{q}$ be the distribution of $G(W)$,
$$
\Pr\{ \| q - \tilde{q} \|_1  \geq \eta + \epsilon \}
\leq 2 |\cT| \exp (- \kappa_0 M \alpha \eta^2)
$$
for some constant $\kappa_0$.
%\item There exists a particular deterministic map
%$G: \{1,...,M\} \rightarrow \cY$ for which
%$$ 
%\| q - \tilde{q} \|_1  \leq \sqrt{\frac{\log 4|\cT|}{\kappa_0M\alpha}} + \epsilon.
%$$
%\end{enumerate}
\end{lemma}

\begin{proof}
Consider the indicator function $I(G(i) = y)$ taking values in $\{0,1\}$. 
Observe that $I(G(i) = y)$ for $i \in \{1,...,M\}$ are i.i.d. random variables 
with expectation value
$
\bbE I(G(i) = y) = q(y).
$
The distribution $\tilde{q}(y)$ of $G(W)$ is $\frac{1}{M} \sum_{i=1}^M I(G(i) = y)$.
By the Chernoff bound (\ref{eq:chernoff}), for each $y\in \cT$, for $\eta \leq \frac{1}{2}$, and some constant $\kappa_0$,
\be
\Pr \left\{ \left|\frac{1}{M} \sum_{i=1}^M I(G(i) = y) - q(y) \right| \geq q(y) \eta \right\}
 \leq 2 \exp (- \kappa_0 M \alpha \eta^2).
\ee
By the union bound,
$$
\Pr\{ {\rm not} \,\,  \iota \} \leq 2 |\cT| \exp (- \kappa_0 M \alpha \eta^2),
$$
where the logic statement $\iota$ is given by
$$
\iota = \left\{ \tilde{q} \in [\hat{q}(1 - \eta), \hat{q}(1 + \eta) ] \right\}
$$
and $\hat{q}(y) = q(y) I(y \in \cT)$.
It remains to relate $\iota$ to a statement about  $\|\tilde{q} - q \|_1$.
First observe that
\be  
\begin{split}
\| \hat{q} - q \|_1 &= \sum_{y} | \hat{q}(y) - q(y) | \\
&= \sum_{y \not\in \cT} q(y)  \leq  \epsilon.
\label{eq:oksd}
\end{split}
\ee
Second, observe that $\iota$ implies 
$
\|\tilde{q} - \hat{q} \|_1 \leq \eta.
$
The two give, via the triangle inequality
$$
 \| q - \tilde{q} \|_1 \leq \eta + \epsilon. 
$$
The statement of the lemma follows.
\end{proof}
\vspace{2mm}

\begin{corollary}
\label{cor:rd}
Consider a random variable $Y$ with distribution $q(y)$, and
let $W$ be the random variable uniformly distributed on $\{1,\ldots,M\}$.
For random variables $Y_1,Y_2,...,Y_M$ all distributed
according to $q^n$, define the map $G: \{1,...,M\} \rightarrow \cY^n$  by $G(i) = Y_i$. 
Let $\tilde{q}$ be the distribution of $G(W)$.
Then, for all $\epsilon, \delta > 0$ and sufficiently large $n$, 
$$
\Pr\{ \| q^n - \tilde{q} \|_1  \geq 2 \epsilon \}
\leq 2 \gamma \exp (- \kappa_0 M \epsilon^2/\gamma),
$$
where $\gamma = 2^{n [H(Y) + c \delta]}$ and $c$ is some positive constant.
\end{corollary}

\begin{proof}
We will assume familiarity with the properties of
typicality and conditional typicality, collected in the Appendix.
We can relate to Lemma $\ref{lemma:rd}$ through the identifications: 
$\cY\ra\cY^n$, $q(y)\ra{q}^n(y^n)$,  and $\cT \ra {\cT}^n_{Y,\delta}$.
The two conditions now read
\begin{align}
q^n(\cT^n_{Y,\delta}) & \geq  1 - \epsilon, \label{cor:rd1}\\
{q^n}(y^n) &\geq  \gamma^{-1}, \,\,\, \forall y^n\in\cT^n_{Y,\delta}.\label{cor:rd2}
\end{align}
These follow from properties $1$ and $2$ of Theorem \ref{a1}
(relabeling $X$ to $Y$ and $p$ to $q$).

\end{proof}

Our next lemma contains the crucial ingredient of the direct coding theorem and is based on
\cite{Winter01a}.
It will tell us how to define the encoding and decoding operations for
a particular value of the common randomness.

\begin{lemma}[Covering lemma]
\label{lemma:c}
We are given a probability distribution $q(y)$ and a conditional probability distribution $P(x|y)$,
with $x\in \cX$ and $y \in \cY$. Assume the existence of sets $\cT \subseteq \cX$ and 
$(\cT_y)_{y\in\cY} \subseteq \cX$ with the 
following properties for all $y\in\cY$:
\begin{align}
\sum_{y\in\cY} q(y)P(\cT_y|y) \;&\geq\; 1-\eps, \label{eq:cov1}\\
\sum_{y\in\cY} q(y)P(\cT|y) \;&\geq\; 1-\eps,\label{eq:cov2}\\
|\cT| \;&\leq\; K,\label{eq:cov3}\\
P(x|y) \;&\leq\; k^{-1}, \quad\quad\forall x\in\cT_y. \label{eq:cov4}
\end{align} 
Define $M=\left\lceil \eta^{-1}K/k\right\rceil$ for some $0<\eta<1$. Given random variables $Y_1,Y_2,...,Y_M$ 
all distributed according to $q$, define the map $D:\{1,2,...,M\}\ra\cY$ by $D(i)=Y_i$. Then there exists a 
conditional probability distribution $E(i|x)$ defined for $i \in \{1,2,...,M\}$ such that 
\be
\Pr\{\|\hat{P} u - Ep \|_1 \geq 5\eps\} \leq 
2K \exp (- \kappa_0\eps^3/\eta),
\label{verysame}
\ee
where $\hat{P}(x|i) = P(x|D(i))$, $u$ is the uniform distribution on $\{1,2,...,M\}$ and $p$ is the marginal distribution defined by $p(x)=\sum_{y\in\cY}P(x|y)q(y)$.
\end{lemma} 

\begin{figure}%[htbp]
	\centerline{ \scalebox{0.65}{	\includegraphics{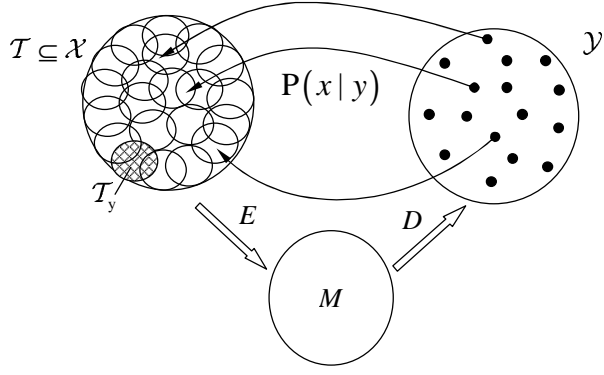}}}	
	\caption{The covering lemma.}
\end{figure}  

\begin{remark}
The meaning of the covering lemma is illustrated in Figure 3.
A uniform distribution on the set $\{1,2,...,M\}$ is diluted via the map $D$ to
the set $\cY$, and then stochastically mapped to the set $\cX$ via $P(x|y)$.
Condition (\ref{verysame}) says that the very same distribution on $\{1,2,...,M\} \times  \cX$ can be 
obtained by starting with the marginal $p(x)$ and stochastically ``concentrating''
it to the set $\{1,2,...,M\}$. For this to be possible, the conditional outputs of the channel 
$P(x|y)$ (for particular values of $y$) 
should be sufficiently spread out to cover the support of $p(x)$.
Each conditional output random variable is supported on $\cT_y$ (\ref{eq:cov1}) 
of cardinality roughly $\geq k$  (\ref{eq:cov4}), and $p(x)$ is supported on $\cT$ (\ref{eq:cov2})
of cardinality $\leq K$  (\ref{eq:cov3}). Thus roughly $M \approx K/k$ conditional random variables
$\hat{P}(x|i)$ should suffice for the covering.
%The covering lemma says that $\|\hat{P} u - Ep \|_1 $ is small. Let $\hat{p}(x) = \sum_i \hat{P}(x|i) u(i)$. Then
%by the properties of trace distance, $\|\hat{p} - p \|_1$ is small. In other words, the distribution $p$ 
%has beed simulated well by covering $\cX$ with the $\cT_{D(i)}$ blobs.
%Note that the packing lemma reduces to the dilution lemma when $P(x|y)$ is the identity channel and
%$\cT_y = \{y\}$. In this case $D$ plays the role of the dilution operation.
\end{remark}

\medskip

\begin{proof}
The idea is to use the Chernoff bound, as in the proof of the randomness dilution lemma. 
First we trim our conditional distributions to make them fit the conditions of the Chernoff bound;
the resulting bound is then related to the condition (\ref{verysame}).

Define 
$$
w(x) = \sum_{y\in\cY} q(y) P(x|y) I(x \in \cA_y),
$$
with $\cA_y=\cT_y\bigcap\cT$ and $\cA=\bigcup_{y\in\cY}\cA_y$.
By properties (\ref{eq:cov1}) and (\ref{eq:cov2}),
$w(\cA) = \sum_{y\in\cY} q(y) P(\cA_y|y)\geq 1-2\eps$. % and hence $w(\cA)\geq 1-2\eps$.
Further define $\cB_y=\cA_y\bigcap\{x:w(x)\geq \eps/K\}$ and $\cB=\bigcup_{y\in\cY}\cB_y$.
Then define 
%we have the \emph{pruned} distribution of $p_y$ and $w$: 
$$
\tilde{P}(x|y) = P(x|y) I(x \in \cB_y), \quad \quad
\tilde{w}(x) = \sum_{y\in\cY}q(y)\tilde{P}(x|y) = w(x) I(w(x)\geq \eps/K).
$$ 
By (\ref{eq:cov3}), the cardinality  of $\cA$ is upper-bounded by $K$, 
those $x\in\cA$ with $w(x)$ smaller than $\eps/K$ contribute at most $\eps$ to 
$w(\cA)$. Thus 
\begin{equation}
\tilde{w}(\cB) \geq w(\cA)-\eps\geq 1-3\eps \label{ineq}.
\end{equation}
Observe 
$$
\bbE \tilde{P}(x|D(i)) = \tilde{w}(x) \geq \epsilon/K.
$$
By  (\ref{eq:cov4}), $0\leq \tilde{P}(x|D(i)) \leq  k^{-1}$.
We can now apply the Chernoff bound (\ref{eq:chernoff}) to the i.i.d. random variables 
$\tilde{P}(x|D(i))$ (for fixed $x \in \cX$) 
\be
\begin{split}
\Pr\left\{\frac{1}{M}\sum^M_{i=1} \tilde{P}(x|D(i)) \notin
 [(1-\eps)\tilde{w}(x),(1+\eps)\tilde{w}(x) ]\right\}
& \leq 2 \exp(-\kappa_0 M \tilde{w}(x) k \eps^2 )  \\
& \leq  2 \exp(-\kappa_0 \eps^3 / \eta ).
\end{split}
\ee
Hence
\be
\Pr\{ {\rm not \,\,} \iota \}
\leq 2K \exp(-\kappa_0 \eps^3/\eta),
\label{eq:jota}
\ee
where the logic statement $\iota$ is defined as
$$
\iota =  \left\{\frac{1}{M}\sum^M_{i=1} \tilde{P}(\cdot|D(i)) \in
 [(1-\eps)\tilde{w}, (1+\eps)\tilde{w}]\right\}.   
$$
%where $M=\left\lceil\eta^{-1}K/k\right\rceil$ for some $0<\eta<1$.
%We see $M$ sets of $\cB(D(i))$ amount to ``cover'' $\cB$, so it must be good for 
%$M$ sets of $\cT_y$ to ``cover'' $\cT$.
%Our covering code $\cC$ is $\{D(i)\}_{i=1,...,M}$ and the covering 
%becomes more and more uniform when $\eta$ tends to 0.
Assume that $\iota$ holds.
%\begin{equation}
%\frac{1}{M}\sum^M_{i=1}\tilde{P}(x|D(i)) \in [(1\pm\eps)\tilde{w}(x)]. 
%\label{conI}
%\end{equation}
Then we can define our conditional distribution $E$ as 
\begin{equation*}
E(i|x)=\frac{1}{(1+\eps)M}\frac{\tilde{P}(x|{D(i)})}{p(x)}.
\end{equation*}
By $\iota$ and the definition of $\tilde{w}$, we can check $E(i|x)$ is a subnormalized conditional 
distribution,
$$\sum^M_{i=1}E(i|x)=\sum^M_{i=1}\frac{1}{(1+\eps)M}\frac{ \tilde{P}(x|D(i))} {p(x)}
\leq\frac{\tilde{w}(x)}{p(x)}\leq 1 .$$

Finally, we estimate $\|\hat{P}u-Ep\|_1$. It is sufficient to
do this for the constructed subnormalized conditional distribution, because we 
can distribute the rest weight to fill up to $1$ arbitrarily. 
The joint distribution of $\hat{P}u$ is
$\{\frac{1}{M} P(x|D(i)) \}$, thus
\be
\|\hat{P}u - Ep\|_1=\sum_{i=1}^M\!\sum_{x\in\cB_{D(i)}}\!\frac{1}{M}
\left(1-\frac{1}{1+\eps}\right) P(x|D(i))
+\sum_{i=1}^M\!\sum_{x\notin\cB_{D(i)}}\!\frac{1}{M} P(x|D(i)).
\label{eq:termz}
\ee
Since $P(\cB_{D(i)}|{D(i)} )\leq 1$, we can bound the first term by $\eps$. 
%Note ($\ref{conI}$) implies $\cB=\bigcup_{i=1}^M\cB_{D(i)}$, i.e. 
%$\frac{1}{M}\sum_{i=1}^M\tilde{p}_{D(i)}\!(x)\in [(1\pm\eps)\tilde{w}(x)]\; \forall x\in\cB$, thus by ($\ref{ineq}$) 
%%we have
By assumption, 
$$
\sum_{x\in\cB}
\frac{1}{M}\sum^M_{i=1}\tilde{P} (x|{D(i)}) \geq 
(1-\eps)\tilde{w}(\cB)\geq 1-4\eps,
$$
Since $\cB_{D(i)} \subseteq \cB$, the second term in (\ref{eq:termz}) 
is bounded by $4 \epsilon$.
%Therefore $\frac{1}{M}\sum_{i=1}^M\sum_{x\notin\cB_{D(i)}} P(x|D(i))\leq 4\eps$.
We have now shown that if $\iota$ holds true then
$$
\|\hat{P}u-Ep\|_1 \leq 5 \epsilon.
$$
Combining with (\ref{eq:jota}) proves the theorem.
\end{proof}
%\vspace{2mm}

\begin{corollary}
\label{cor:c}
Consider the joint random variable $XY$ distributed according to $q(y) P(x|y)$.
Given random variables $Y_1,Y_2,...,Y_M$ 
all distributed according to $q^n$, define the map $D:\{1,2,...,M\}\ra\cY^n$ by $D(i)=Y_i$. 
Then, for all $\epsilon, \delta > 0$ and sufficiently large $n$, there exists a 
conditional probability distribution $E(i|x^n)$ defined for $i \in \{1,2,...,M\}$ such that 
\be
\Pr\{\|\hat{P} u - E p^n \|_1 \geq 5\eps\} \leq 
2\alpha \exp (- \kappa_0M\eps^3 \beta/\alpha),
\ee
where $\hat{P}(x^n|i) = P^n(x^n|D(i))$, $u$ is the uniform distribution on $\{1,2,...,M\}$, $p$ is 
the marginal distribution defined by $p(x)=\sum_{y\in\cY}P(x|y)q(y)$, $\alpha = 2^{n[H(X)+c\delta]}$,
$\beta=2^{n[H(X|Y)-c\delta]}$.
\end{corollary}

\begin{proof}
We can relate to Lemma \ref{lemma:c} through the identifications (see Appendix): 
$\cX\ra\cX^n$, $\cY\ra \cY^n$, $q(y)\ra{q}^n(y^n)$, 
$P(x|y)\ra{P}^n(x^n|y^n)$, $\cT\ra\cT^n_{X,\,3\delta}$, and 
$\cT_y\ra \hat{\cT}^n_{X|Y,\,\delta}(y^n)$, with
$$
\hat{\cT}^n_{X|Y, \delta}(y^n)= \left\{\begin{array}{cc} 
{\cT}^n_{X|Y, \delta}(y^n) \,\,  & y^n \in \cT^n_{Y, \delta} \\
\emptyset \,\, & {\rm otherwise.}
\end{array} \right.
$$

The four conditions now read (for all $y^n\in\cY^n$),
\begin{align}
\sum_{y^n\in\cY^n} q^n(y^n)P^n(\hat{\cT}^n_{X|Y,\,\delta}(y^n)|y^n) & \geq 1- 2\eps, \label{cor:c1}\\
\sum_{y^n\in\cY^n} q^n(y^n)P^n(\cT^n_{X,\,3\delta}|y^n) & \geq 1- 2\eps, \label{cor:c2}\\
|\cT^n_{X,\,3\delta}| & \leq \alpha, \label{cor:c3}\\
P^n(x^n|y^n) & \leq \beta^{-1},\quad\quad\forall x^n\in\hat{\cT}^n_{X|Y, \delta}(y^n). \label{cor:c4}
\end{align}
These follow from Theorem \ref{a2}, 
switching the roles of $X$ and $Y$ and setting $\delta = \delta'$.
\end{proof}

\medskip

We will also need the Holevo-Schumacher-Westmoreland (HSW) theorem
\cite{Holevo98, SW97}.

\begin{proposition}[HSW Theorem]
\label{prop:HSW}
Given an ensemble 
$$
\sigma^{YB} = \sum_{y \in \cY} q(y) \proj{y}^Y \otimes {\bar \rho}_y^B,
$$
and integer $n$, consider the encoding map 
$F:\{0,1\}^{nS} \rightarrow \cY^n$ given by $F(s) = Y_s$,
where  the $\{Y_s \}$ are random variables chosen according
to the i.i.d. distribution $q^n$.
For any $\epsilon, \delta>0$ and sufficiently large $n$,
there exists a decoding POVM $\{ \Lambda_s \}_{s \in \{0,1\}^{nS}}$
on $B^n$ for the encoding map $F$ with  $S = I(Y;B)_\sigma - \delta$,
such that for all $s$,
$$\bbE \sum_{s'} | \pi({s'|s})  - \delta(s,s') | \leq \epsilon.$$
Here $\pi({s'|s})$ is the probability of decoding $s'$ conditioned on $s$ having
been encoded:
\be
 \pi({s'|s}) = \tr( \Lambda_{s'}  {\bar \rho}_{F(s)}),
\label{191/2}
\ee
$\delta(s,s')$ is the delta function
and the expectation is taken over the random encoding.
\end{proposition}

Now we are ready to prove the direct coding theorem:
\vspace{1mm}

\noindent {\bf{Proof of Theorem 1 (direct coding)}} \space \space  
%Our approach is to construct simulation codes, each
%of which consists of HSW codes and can ``cover'' the typical set $\cT^n_{X,\,3\delta}$ well, 
%and by the covering lemma and HSW theorem~\cite{Holevo98, SW97} we can see how to encode input 
%and simulate output with the help of common randomness and quantum side information explicitly.
%A graphical representatoin of the proof is given in the Figure 2.
Fix $\epsilon, \delta > 0$ and a sufficiently large $n$ (cf. Corollaries \ref{cor:rd},
\ref{cor:c} and Proposition \ref{prop:HSW}).
Consider the random variables $Y_{lms}$,
$l\in\{0,1\}^{nC}$, $m\in\{0,1\}^{nR}$, $s=\{0,1\}^{nS}$ (for some $C,R$ and $S$ to be specified
later), independently distributed according to $q^n$, where 
$q(y) = \sum_x p(x) W(y|x)$. The $Y_{lms}$ are going to
serve simultaneously as a ``randomness dilution code'' $G(l,m,s) = Y_{lms}$
(cf. the $Y_1, \dots, Y_M$ in Corollary \ref{cor:rd}, $M$ here being $2^{n (C + R + S}$);
as $2^{nC}$ independent ``covering codes'' $D_l(m,s) = Y_{lms}$ (cf. the 
$Y_1, \dots, Y_M$ in Corollary \ref{cor:c}, $M$ here being $2^{n (R + S}$);
and as $2^{n(C + R)}$ independent HSW codes $F_{lm}(s) = Y_{lms}$ (cf. Proposition \ref{prop:HSW}).
We will conclude the proof by ``derandomizing'' the code, i.e. showing that
a particular realization of the random $Y_{lms}$ exists with suitable properties.
% which accomplishes our goal.

Define, as in the two corollaries, $\alpha = 2^{n[H(X) + c \delta]}$,
 $\beta = 2^{n[H(X|Y) - c \delta]}$, and $\gamma = 2^{n[H(Y) + c \delta]}$.
Define two independent uniform distributions $u'(l)$ and $u(ms)$ on the sets $\{0,1\}^{nC}$ and 
$\{0,1\}^{nR} \times \{0,1\}^{nS}$,
respectively.
The stochastic map $\tilde{D}(y^n|l,m,s)$ is defined as
$$
\tilde{D}(y^n|l,m,s) = I(y^n = D_l(m,s)).
$$
Corollary \ref{cor:c} defines corresponding encoding stochastic maps
$\{E_l(m,s|x^n)\}$.
For any 
$l\in\{0,1\}^{nC}$, define the logic statement $\iota_l$ by
$\xi_l \leq 5 \epsilon$, where
$$ 
\xi_l = 
\sum_{m,s}\sum_{x^n}\left|\sum_{y^n}P^n(x^n|y^n)\tilde{D}
(y^n|l,m,s)u(ms)-E_l(m,s|x^n)p^n(x^n)\right|.  
$$
By Corollary \ref{cor:c}, for all $l$
\begin{equation}
\Pr\{{\rm not}\,\,\iota_l\}\leq  2\alpha \exp(-2^{n(R+S)} \kappa_0\eps^3\beta/\alpha).
\label{p1}
\end{equation}
Define the logic statement $\iota'$ by $\xi' \leq 2 \epsilon$, where
$$
\xi' = \sum_{y^n} \left| \sum_{l,m,s}\tilde{D}(y^n|l,m,s)u'(l)u(ms) - 
q^n(y^n)  \right|.
$$
By Corollary \ref{cor:rd},
\be
\Pr\{{\rm not}\,\,\iota'\}\leq 2 \gamma \exp(-2^{n(C+R+S)} \kappa_0 \eps^2/ \gamma).
\label{p2}
\ee
%Thus we have totally $L$ sets of simulation codes $(S_l)_{l=1,2,...,L}$, each of which has 
%$MS$ elements and can cover the set $\cT^n_{X,\,\delta}$ well. 
Once we fix the randomness we shall be using 
\be
\tilde{W}(y^n|x^n)=\sum_{l,m,s}\tilde{D}(y^n|l,m,s)E_l(m,s|x^n)u'(l)
\ee
to simulate the channel $W^n(y^n|x^n)$.
Observe that
\begin{align}
& \sum_{x^n y^n} \left| p^n(x^n) (W^n(y^n|x^n) - \tilde{W}(y^n|x^n)) \right| \label{distance}\\
 %\|p^n\;\tilde{W} - p^n\;W^n\|_1
&=\sum_{x^n,y^n}\left|\sum_{l,m,s}\tilde{D}(y^n|l,m,s)E_l(m,s|x^n)u'(l)p^n(x^n)
-W^n(y^n|x^n)p^n(x^n)\right| \nonumber \\
&\leq\sum_{x^n,y^n}\sum_{l,m,s}\tilde{D}(y^n|l,m,s)u'(l)\left|E_l(m,s|x^n)p^n(x^n)-
\!\!\sum_{\hat{y}^n}P^n(x^n|\hat{y}^n)\tilde{D}(\hat{y}^n|l,m,s)u(ms)\right|\nonumber  
\\&+\sum_{x^n,y^n}P^n(x^n|y^n)
\left|\sum_{l,m,s}\tilde{D}(y^n|l,m,s)u'(l)u(ms) - q^n(y^n)\right|\nonumber  \\
&\leq \max_l \xi_l + \xi'. \label{xixi}
\end{align}
To obtain the first inequality we have used
$$
\tilde{D}(y^n|l,m,s)\tilde{D}(\hat{y}^n|l,m,s)=\tilde{D}(y^n|l,m,s)\delta(y^n,\hat{y}^n)
$$
and the triangle inequality.
%In the last line, the first term is bounded accordng to $\iota_l$ and
%the second term  according to $\iota'$.

We shall now invoke Proposition \ref{prop:HSW}.
Define $q(y) {\bar \rho}_y = \sum_x p(x) W(y|x) \rho_x$.
Setting $F_{lm}(s) = Y_{lms}$ and $S =  I(Y; B)-c\delta$,
there exists a set $\{\Lambda^{(lm)}\}_{lm\in\{0,1\}^{n (C+R)}}$, where each 
$\Lambda^{(lm)}=\{\Lambda^{(lm)}_{s'}\}_{s'\in\{0,1\}^{n S}}$ is a POVM acting on $B^n$, 
such that
\be
\bbE \sum_{s'} | \pi_{lm}({s'|s})  - \delta(s,s') | \leq \epsilon
\label{hswsw}
\ee
for all
$l,m$ and $s$.
$ \pi_{lm}({s'|s}) $ describes the noise experienced in conveying $s$ to Bob,
if the channel $W^n(y^n|x^n)$ were implemented exactly.
However, Alice only has the simulation $\tilde{W}(y^n|x^n)$, which
corresponds to the ensemble
$ \tilde{q}(y^n) \tilde{\rho}_{y^n} := 
\sum_{x^n} p^n(x^n)\tilde{W}(y^n|x^n) \rho_{x^n}$.

Observe that (\ref{distance}) is another way of expressing 
$||(\rho^{X\bar{Y}B})^{\otimes n} - \sigma^{X^n\tilde{Y}^nB^n}||_1 = 
||(\rho^{X\bar{Y}})^{\otimes n} - \sigma^{X^n\tilde{Y}^n}||_1$.
Applying monotonicity of trace distance to (\ref{xixi}), we have
$$
||(\rho^{\bar{Y}B})^{\otimes n} - \sigma^{\tilde{Y}^nB^n}||_1 = \sum_{y^n} \| 
q^n(y^n)  {\bar \rho}_{y^n} - \tilde{q}(y^n) \tilde{\rho}_{y^n} \|_1
\leq \max_l \xi_l + \xi',
$$
and hence by the triangle inequality and monotonicity of trace distance
$$
 \bbE \|{\bar \rho_{F(s)} - \tilde{\rho}_{F(s)}}\|_1 \leq \sum_{y^n}
 ||q^n(y^n)\bar{\rho}_{y^n} - \tilde{q}(y^n)\tilde{\rho}_{y^n}||_1 + \sum_{y^n}|\tilde{q}(y^n) - q^n(y^n)| \leq 2 (\max_l \xi_l + \xi').
$$
Thus, the actual noise experienced in conveying $s$ to Bob, denoted  by 
$\tilde{\pi}_{lm}({s'|s})$, obeys
$\bbE \sum_{s'} | \pi_{lm}({s'|s})  -  \tilde{\pi}_{lm}({s'|s})| \leq  
2 (\max_l \xi_l + \xi')$. Combining the above with (\ref{hswsw}) gives
$$
\bbE \sum_{s'} | \tilde{\pi}_{lm}({s'|s})  - \delta(s,s') | \leq  
2 (\max_l \xi_l + \xi') + \epsilon.
$$
Let us focus on the effect this imperfection in the HSW decoding will have on the simulation.
By monotonicity,
$$
\bbE \sum_{x^n \tilde{y}^n y^n} 
|\sum_{l,m,s,s'}  \tilde{D}(y^n|lms) \tilde{D}(\tilde{y}^n|lms') 
E_l(ms|x^n)u'(l)  p^n(x^n) (\tilde{\pi}_{lm}({s'|s})  - \delta(s,s')) | \leq 
2 (\max_l \xi_l + \xi') + \epsilon.
$$
By the Markov inequality, $\Pr\{{\rm not}\,\,\iota''\}\leq  \frac{1}{2}$,
where $\iota''$ is the logic statement 
$$
\sum_{x^n \tilde{y}^n y^n} 
\left|\sum_{l,m,s,s'}  \tilde{D}(y^n|lms) \tilde{D}(\tilde{y}^n|lms') 
E_l(ms|x^n) u'(l) p^n(x^n) (\tilde{\pi}_{lm}({s'|s})  - \delta(s,s')) \right| \leq 
4 (\max_l \xi_l + \xi') + 2 \epsilon.
$$ 

Now for the derandomization step. Pick $C= H(Y|X)-c\delta$ and $R = I(X;Y)-I(Y;B)+4c\delta$. 
By the union bound $\iota_l$ for all $l$, $\iota'$, and $\iota''$ hold true with 
probability $>0$.
Hence there exists a specific choice of $\{ Y_{lms} \}$ for which 
all these conditions are satisfied. Consequently,
$$
\sum_{x^n \tilde{y}^n y^n} 
\left|\sum_{l,m,s,s'}  \tilde{D}(y^n|lms) \tilde{D}(\tilde{y}^n|lms') 
E_l(ms|x^n) u'(l) p^n(x^n) (\tilde{\pi}_{lm}({s'|s})  - \delta(s,s')) \right| \leq 30 \epsilon,
$$
i.e. $||\sigma^{X^n\tilde{Y}_o^n\tilde{Y}^n} - \sigma^{X^n\hat{Y}^n\tilde{Y}^n}||_1\leq 30\eps$,
where $\tilde{Y}_o^n = \tilde{Y}^n$ is Bob's simulation output random variable if his decoding measurement is perfect.
Combining with (\ref{xixi}) ($||(\rho^{XY\bar{Y}})^{\otimes n} - \sigma^{X^n\tilde{Y}_o^n\tilde{Y}^n}||_1 
\leq 7\eps$) gives
$$
\|(\rho^{XY\bar{Y}})^{\otimes n} - \sigma^{X^n\hat{Y}^n\tilde{Y}^n} \|_1 \leq
37 \eps.
$$
This is almost what we need.
The statement of the theorem also insists that the state of the $B^n$ system
is not much perturbed by the measurement. The crucial ingredient ensuring this,
as in \cite{DW02},
is the gentle measurement lemma \cite{Winter99}. To improve readability,
we omit the details of its application here.
%In conclusion, there exists an $(n,R+\delta, C+\delta, \eps)$ code. 
%This proves the direct coding theorem.
\qed 
%Then we will show that our code construction above is optimal under the hypothesis that the channel with 
%asymptotically noiseless feedback is simulated. 

\medskip

Before proving the converse, recall Fannes' inequality~\cite{Fannes73}:
\begin{lemma}[Fannes' inequality]
Let $P$ and $Q$ be probability distributions on a set with finite cardinality $d$, 
such that $\|P-Q\|_1 \leq \eps$. Then
  $\bigl|H(P)-H(Q)\bigr| \leq \eps\log d + \tau(\eps)$, with
  \begin{equation*}
    \tau(\eps) = \begin{cases}
                     -\eps\log\eps   & \text{ if } \eps\leq 1/4, \\
                     1/2               & \text{ otherwise.}
                   \end{cases}
  \end{equation*}
  Note that $\tau$ is a monotone and concave function and $\tau(\eps)\ra 0$ as $\eps\ra 0$.
  \qed
\end{lemma}

\noindent {\bf{Proof of Theorem 5 (converse)}} \space \space
Consider an $(n,R,C,\eps)$ code. Define the uniform random variable $U$ on the set $\{0,1\}^{nC}$ to denote
 the common randomness, and  $W$ on the set
 $\{0,1\}^{nR}$ to denote the encoded message sent to Bob. 
We have the following Markov chain
$$X^n\!\rightarrow B^nWU\!\rightarrow \hat{B}^n\hat{Y}^n.$$
%where $\tilde{B}^n$ is the new label for Bob's post-measurement system.
The following chain of inequalities  holds:
\begin{align*}
n R   &\geq H(W|U) \\
     &= H(W|U) + I(X^n; B^n|U) - I(X^n; B^n) \\
%     &= I(X^n; B^nW|U) + I(W; B^n|U) + H(W|X^nB^nU) - I(X^n; B^n) \\
     &\geq I(X^n; B^nW|U) - I(X^n; B^n) \\
  & =   I(X^n; B^nW U)  - I(X^n; B^n) \\
  &\geq I(X^n; \hat{B}^n\hat{Y}^n) - I(X^n; B^n) \\
&\geq n\left(I(X; BY) - I(X; B) - f(n,\eps)\right) \\
& = n\left(I(X; Y) - I(Y; B) - f(n,\eps)\right).
\end{align*}
with $f(n,\eps)\ra 0$ as $n\ra\infty$ and $\eps\ra 0$.
%$(a)$ follows trivially from  $W \in \{0,1\}^{n R}$, and 
The second line  from $I(X^n; B^n|U)=I(X^n; B^n)$, 
%$(b)$ comes from the non-negativity of conditional mutual information and conditional entropy. 
and the  fourth from $I(X^n; U) = 0$. 
The fifth line  is the data processing inequality based on the Markov chain above.
The sixth is a consequence of Fannes inequality, and the last line is 
based on the Markov chain $Y^n\!\rightarrow X^n\!\rightarrow B^n$. 
%such that 
%\begin{align*}
%I(X^n; B^n) = I(X^nY^n; B^n)= I(X^n; B^nY^n) - I(X^n; Y^n) + I(Y^n; B^n) . 
%\end{align*}

Based on the Markov chain 
$$\tilde{Y}^n\!\rightarrow B^nWU\!\rightarrow \hat{Y}^n,$$
 we have another chain of inequalities :
\begin{align*}
n R + n C &\geq H(W) + H(U) \\
          &\geq H(WU) \\
%          &= H(WU) + I(\tilde{Y}^n; B^n) - I(\tilde{Y}^n; B^n) \\
          &= I(\tilde{Y}^n; B^nWU) + I(WU; B^n) + H(WU|\tilde{Y}^nB^n) - I(\tilde{Y}^n; B^n)\\
          &\geq I(\tilde{Y}^n; B^nWU) - I(\tilde{Y}^n; B^n) \\
          &\geq I(\tilde{Y}^n; \hat{Y}^n) - I(\tilde{Y}^n; B^n) \\
          &\geq n(H(Y) - I(Y; B) - f'(n,\eps)) 
\end{align*}
with $f'(n,\eps)\ra 0$ as $n\ra\infty$ and $\eps\ra 0$.
%$(a')$ follows from $W \in \{0,1\}^{n R}$ and $H(U) = nC$. $(b')$ comes from 
%the non-negativity of conditional mutual information and conditional entropy. 
The last two inequalities are from the data processing inequality and Fannes inequality. 
Thus any achievable rate pair $(R,C)$ must obey the conditions of Theorem \ref{thm1}.
%So if we choose $\delta= \max\{f(n,\eps), f'(n,\eps)\}$, we can have $R \geq I(X; Y) - I(Y; B) -\delta$ and $C\geq H(Y|B) -\delta$, as claimed.

\qed
\vspace{1mm}

We can use the theory of resource inequalities~\cite{DHW05} 
to succinctly express our main result. In this case we need to introduce
an additional protagonist, the Source, which starts the
protocol by distributing the state 
$$
\rho^{X_{S}S} = \sum_{x}p (x) \proj{x}^{X_{S}}\otimes \rho_{x}^S,
$$
between Alice and Bob. Alice gets $X_S$ through the classical identity channel
$\bar{\text{id}}^{X_S\ra X_A}$ and Bob gets $S$ through the quantum identity
channel $\text{id}^{S\ra B}$.
The goal is for Alice and Bob to end up sharing the state 
\be
\sigma^{X_A Y_A Y_B B} = \sum_x p(x) \sum_y W(y|x) \proj{y}^{Y_A} \otimes \proj{y}^{Y_B} 
\otimes  \proj{x}^{X_A} \otimes  \rho_{x}^B,
\label{output}
\ee
\emph{as if} $\rho^{X_{S}S}$ was sent through the channel
$W^{X_S \rightarrow Y_A Y_B} \otimes \text{id}^{S\ra B}$ 
(the former is a feedback version of $W$). 
Our direct coding theorem is equivalent to the resource inequality
\be
\begin{split}
\langle\bar{\text{id}}^{X_S\ra X_A}\!\otimes\text{id}^{S\ra B}\!:\!\rho^{X_SS}\rangle
& +(I(X_A;Y_B)_{\sigma}-I(Y_B;B)_{\sigma})[c\ra c]+H(Y_B|X_A)_{\sigma}[c \, c] \\
&\stackrel{s}{\geq} 
\langle W^{X_S\ra Y_A Y_B}\!\otimes\text{id}^{S\ra B}\!:\!\rho^{X_SS}\rangle .
\label{glava}
\end{split}
\ee
The superscript $s$ stands for ``source'' and is a technical subtlety \cite{DHW05}.

%%%%%%%%%%%%%%%%%%%%%%%%%%%%%%%%%%%%%%%%%%%%%%%%%%%%%%%%%%%%%%%%%%%%%%%%%%%%%%%%%%%%%%%%%%%%%%%%%%%%%%%%%%%%%%%%%%%%%%%%
%%%%%%%%%%%%%%%%%%%%%%%%%%%%%%%%%%%%%%%%%%%%%%%%%%%%%%%%%%%%%%%%%%%%%%%%%%%%%%%%%%%%%%%%%%%%%%%%%%%%%%%%%%%%%%%%%%%%%%%%%%%%%%%%%%%%%%%%%%%%%%%%%%%%%%%%%%%%%%%%%%%%%%%%%%%%%%%%%%%%%%%%%%%%%%%%%%%%%%%%%%%%%%%%%%%%%%%%%%%%%%%%%%%%%%%%%%%%%%%%%
\section{Applications}
In this section, common randomness distillation %from bipartite classical-quantum states 
and rate-distortion coding with side information will be seen as simple corollaries
of our main result. %Simple proof of both the direct coding theorems can be made by applying our result.

\subsection{Common randomness distillation}
Alice and Bob share $n$ copies of a bipartite classical-quantum state 
$$
\rho^{X_{A}B} = \sum_x p(x) \ket{x}\bra{x}^{X_{A}}\otimes\rho^B_{x} ,
$$
 and Alice is allowed a rate $R$ bits of classical communication to Bob. 
Their goal is to distill a rate $C$ of common randomness (CR).
In terms of resource inequalities, a CR-rate pair $(C,R)$ is said to be
achievable iff 
$$
\langle\rho^{X_{A}B}\rangle + R \, [c\ra c] \geq C \, [c \,c] .
$$
Define the CR-rate function $C(R)$ to be 
$$
C(R) = \sup\{C : (C, R)\; \text{is achievable}\}.
$$
and the distillable CR function as $D(R) = C(R) - R$.
The following theorem was proved in \cite{DW03a}.
\begin{theorem}%[\cite{DW03a}]
Given the classical-quantum system $XB$, then
$$
 D(R) = \max_{Y|X}\{I(Y; B)~ |~ I(X; Y) - I(Y; B)\leq R\}.
$$
where $C(R) = C^*(R) = R + D^*(R)$. The maximum is over all conditional probability distributions
$W(y|x)$ with $|\cY|\leq|\cX|+1$. 
\end{theorem}

We give below a concise proof of the direct coding part of this theorem, relying
on our main result (\ref{glava}) and the resource calculus \cite{DHW05}. 
%The proof is similar to the proof of $CR$ distillation in~\cite{DHW05} by 
%using channel simulation with quantum side information instead of classical compression 
%with quantum side information.

\medskip

\begin{proof}
%Since we can simulate a channel with asymptotically noiseless feedback, we can do 
%common randomness ($CR$)
%concentration based on $\tilde{Y}\hat{Y}$. This assures us 
%to distill common randomness from bipartite state $XB$ via channel simulation and 
%common randomness concentration~\cite{DHW05}.
%To be succinct, we will rephrase and prove the direct coding theorem in terms of 
%resource inequalities. 
We need to prove 
\begin{equation}
\langle\rho^{X_{A}B}\rangle + (I(X_A; Y_B)_{\sigma} - I(Y_B; B)_{\sigma}) [c \ra c] 
\geq I(X_A; Y_B)_{\sigma} [c \, c] \label{disccout} ,
\end{equation}
with  $\sigma^{X_A Y_A Y_B B}$ given by (\ref{output}).
Observe the following string of resource inequalities:
\begin{align*}
\quad& \langle\bar{\text{id}}^{X_S\ra X_A}\otimes\text{id}^{S\ra B}: \rho^{X_SS}\rangle
 + (I(X_A; Y_B)_{\sigma} - I(Y_B; B)_{\sigma}) [c\ra c] + H(Y_B|X_A)_{\sigma} [c \, c] \\
& \geq \langle W^{X_S\ra Y_AY_B}\otimes\text{id}^{S\ra B}: \rho^{X_SS}\rangle \\
& \geq \langle W^{X_S\ra Y_AY_B} : \rho^{X_S}\rangle \\
& \geq \langle W^{X_S\ra Y_AY_B}(\rho^{X_S})\rangle \\
& \geq H(Y_B)_{\sigma} [c \, c] .
\end{align*}
The first inequality is by (\ref{glava}) and Lemma 4.11 of \cite{DHW05} which
allows us to drop the $s$ superscript; 
the second and third are by parts $5$ and $2$, respectively, 
of Lemma $4.1$ of~\cite{DHW05}. The last inequality is common randomness concentration
\cite{DHW05}, which states that $\< \sigma^{Y_A Y_B} \> \geq H(Y_B)_{\sigma} \, [c \, c]$.
By Lemma $4.10$ of~\cite{DHW05}, 
$\langle\bar{\text{id}}^{X_S\ra X_A}\!\otimes\text{id}^{S\ra B}\!:\!\rho^{X_SS}\rangle$
 can be replaced by 
\begin{equation}
\langle\rho^{X_{A}B}\rangle = \langle\bar{\text{id}}^{X_S\ra X_A}\otimes\text{id}^{S\ra B}
(\rho^{X_SS})\rangle \label{fake}. 
\end{equation}
Thus by (\ref{fake}) and Lemma $4.6$ of~\cite{DHW05}, we have
$$
\langle\rho^{X_{A}B}\rangle + (I(X_A; Y_B)_{\sigma} - I(Y_B; B)_{\sigma}) [c\ra c] + o [c \, c]
\geq I(X_A; Y_B)_{\sigma} [c \, c] .
$$
Since $[c \ra c] \geq [c \, c]$, by Lemma $4.5$ of~\cite{DHW05} the $o$ term can be dropped, and
(\ref{disccout}) is proved.
\end{proof}

\subsection{Rate-distortion trade-off with quantum side information}
Rate-distortion theory, or lossy source coding, is a major subfield of 
classical information theory \cite{Berger71}. When insufficient storage space
is available, one has to compress a source beyond the Shannon entropy.
By the converse to Shannon's compression theorem, this means that 
the reproduction of the source (after compression and decompression)
suffers a certain amount of distortion compared to the original. 
The goal of rate-distortion theory is
to minimize a suitably defined distortion measure for a given desired compression rate.
Formally, a distortion measure is a mapping $d: \cX\times{\cX} \ra \bbR^{+}$
from the set of source-reproduction 
alphabet pairs into the set of non-negative real numbers. This function
can be extended to sequences $\cX^n\times {\cX}^n$ by letting 
$$d(x^n, \hat{x}^n) = \frac{1}{n}\sum_{i=1}^nd(x_i, \hat{x}_i).$$

We consider here a quantum generalization of the classical Wyner-Ziv~\cite{WZ76} problem.
The encoder Alice and decoder Bob share $n$ copies of the classical-quantum system $XB$
in the state (\ref{eq:stayte}). Alice sends Bob a classical message at rate $R$, based on
which, and with the help of his side information $B^n$, Bob needs to reproduce $x^n$ with lowest
possible distortion.
An $(n,R,d)$ rate-distortion code is given by an encoding map $\cE_n: \cX^n \rightarrow \{0,1\}^{nR}$
and a decoding map $\cD_n$ which takes $\cE_n(x^n)$ and the state $\rho_{x^n}$ as
inputs and outputs a string $\hat{x}^n \in \cX^n$. $\cD_n$ is implemented by performing a $\cE_n(x^n)$-dependent
measurement, followed by a function mapping $\cE_n(x^n)$ and the measurement outcome to $\hat{x}^n$.
The condition on the reproduction quality is 
%given $n$ copies of a classical-quantum system $XB$ (source $X$ with quantum side information $B$)
%in the state $\rho^{X^nB^n}$, construct an $n$-block code $(\cE_n, \cD_n)$ such that for a given distortion $d$,
$$
d(\cE_n, \cD_n) := \bbE d(X^n, \hat{X}^n) = \sum_{x^n} p^n(x^n)d(x^n, \cD_n(\cE_n(x^n), \rho_{x^n})) \leq d ~.
$$
A pair $(R, d)$ is achievable if there exists an $(n,R+\delta,d)$ code for any $\delta > 0$ and
sufficiently large $n$. Define $R_B(d)$ to be the infimum of rates $R$ for which $(R, d)$ is achievable.
% Define $R^{(n)}_B(d)$ as the minimum 
%$R$ such that $(R,d)$ is achievable if the side information $B$ available to the decoder. 
%The rate-distortion function with quantum side information $R_B(d)$ is defined as

\begin{theorem}
Given $n$ copies of a classical-quantum system $XB$ in the state $\rho^{X^nB^n}$, then
$$
R_B(d) = \lim_{n \ra \infty} R^{(n)}_B(d),
$$
$$
R^{(n)}_B(d) = \frac{1}{n}\min_{Y|X^n}\min_{\cD:YB^n \ra \hat{X}^n} (I(X^n; Y) - I(Y; B^n))
$$
where the minimization is over all conditional probability distributions $W(y|x^n)$, %$|\cY|\leq |\cX|^n + 1$,
 and decoding maps $\cD:YB^n \ra \hat{X}^n$, such that
$$
\bbE{d(X^n, \cD(Y, B^n))} = \sum_{x^n,y}p^n(x^n)W(y|x^n)d(x^n, \cD(y, \rho_{x^n}^{B^n})) \leq d . 
$$
\end{theorem}

Note that $(m+n)R^{(m+n)}_B(d) \leq mR^{(m)}_B(d) + nR^{(n)}_B(d)$. By arguments similar to 
those for the channel capacity (see e.g.~\cite{BNS98}, Appendix A), the limit $R_B(d)$ exists. 
However, the formula of $R_B^{(n)}(d)$ is a ``regularized'' form, so $R_B(d)$ can not be effectively 
computed.

We omit the easy proof of the converse theorem. % is easy to prove and is omitted.
The direct coding theorem is an immediate consequence of Theorem \ref{thm1} (cf. \cite{Winter02a}):

\vspace{2mm}

\noindent  {\bf Proof of Theorem 4.2 (direct coding) }   \space \space
It suffices to prove the  achievability of $R^{(1)}_B(d)$, for a fixed channel $W(y|x)$
and decoding map $\cD: YB \rightarrow \hat{X}$. 
Consider an $(n,R,C,\epsilon)$ simulation code for the channel $W(y|x)$. 
The simulated state $\sigma^{X^n \hat{Y}^n \hat{B}^n}$ can be 
written as a convex combination of simulations corresponding to particular 
values of the common randomness $l$:
$$
\sigma^{X^n \hat{Y}^n \hat{B}^n} = \sum_l u'(l) \sigma_l^{X^n \hat{Y}^n \hat{B}^n}.
$$
In other words, $\sigma_l^{X^n \tilde{Y}^n \hat{B}^n}$ is obtained from the 
encoding $E_l(m,s|x^n)$, POVM set $\{ \Lambda^{(lm)}\}_{ m \in\{0,1\}^{nC}} $, 
and decoding $D_l(m,s)$. 
From the condition for successful simulation (\ref{41/2}) and monotonicity of trace distance it
follows that 
\be
\| \sum_l u'(l) \, \cD^{\otimes n}(\sigma_l^{\hat{Y}^n \hat{B}^n})  - \cD^{\otimes n}(\rho^{YB})^{\otimes n} \|_1
\leq \epsilon.
\label{inv}
\ee
%By Theorem 3.1, we can approximately simulate the state $\rho^{X^nYB^n}$ by using some 
%common randomness $l$ and a deterministic code $(E_{nl}, D_{nl})$ sending $\approx I(X^n;Y)-I(Y;B^n)$ bits. 
For each $l$ define rate-distortion encoding  
$\cE^l_{n}$ by $E_l(m,s|x^n)$, and decoding $\cD^l_n$ by the
 POVM set $\{ \Lambda^{(lm)}\}_{ m \in\{0,1\}^{nC}}$ followed by $D_l(m,s')$ ($s'$ is the POVM
outcome) and $\cD^{\otimes n}$.
Invoking  (\ref{inv}), $\bbE{d(X, \cD(Y, B))} \leq d$  and the linearity of the distortion measure, gives
$$
\sum_{l}u'(l) \, d(\cE_n^{l}, \cD_n^{l}) \leq d +  c_0 \eps,
$$  
for some constant $c_0$. Hence there exists a particular $l$ for which
$$
d(\cE_n^{l}, \cD_n^{l}) \leq d +  c_0 \eps.
$$ 
The direct coding theorem now follows from the achievable rates given by Theorem  \ref{thm1}.
%Minimizing the rate $R$ over all channels $W(y|x^n)$ and corresponding decoding maps $\cD$, 
%$R^{(n)}_B(d)$ is the minimum rate we can achieve within distortion $d$.
\qed

\vspace{2mm}

%\noindent  {\bf Proof of Theorem 4.2 (converse) }   \space \space For any rate-distortion code
%with quantum side information, define the encoding map $\cE_n : X^n \ra Y$
%and the decoding map $\cD_n : YB^n \ra \hat{X}^n$. 
%Let $ Y = \cE_n(X^n)\in \{0,1\}^{nR}$ denote the encoded version of $X^n$. 
%For any conditional distribution $W(y|x^n)$ and decoding map $\cD_n$ satisfying the distortion 
%constraint $\bbE{d(X^n, \cD_n(Y, B^n))} \leq d$, we have the following chain of inequalities
%\begin{align*}
%nR &\geq H(Y) \\
%   &\geq H(Y|B^n) \\
%   &\geq I(X^n ; Y | B^n) \\
%   &= I(X^nB^n ; Y ) - I(Y ; B^n) \\
%   &= I(X^n ; Y) - I(Y ; B^n) \\
%   &\geq nR^{(n)}_B(d) .
%\end{align*}
%The second line comes from the non-negativity of quantum mutual information. The third 
% and the fifth line are based on the Markov chain $Y \ra X^n \ra B^n$ such that 
% $H(Y|X^nB^n) = H(Y|X^n)\geq 0$ and $I(B^n ; Y | X^n) = 0$.
%\qed
%\vspace{2mm}
The classical Wyner-Ziv problem is recovered by making $B$ into a classical system $Z$, i.e.
by setting $\rho_x=\sum_{z}p(z|x)\ket{z}\!\bra{z}$ with $\sum_{z}p(z|x)=1$ and 
associating the joint distribution $p(x) p(z|x)$ with the random variable $XZ$.
% be distributed %, we can make our side 
%information to be a classical random variable $Z$ correlated to the source $X$ with a joint 
%distribution $p(x,z)$. 
In this case a single-letter formula is obtained 
%Then the problem comes back to the classical Wyner-Ziv problem in
%which the formula can be ``single-letterized'', i.e.
$$
R_Z(d) = R^{(1)}_Z(d) = \min_{Y|X}\min_{D : YZ \ra \hat{X}} (I(X ; Y) - I(Y ; Z))~.
$$ 
It is an open question whether a single-letter formula exists for $R_B(d)$.
Following the standard converse proof of ~\cite{CT91, WZ76} we are able to produce
a single letter lower bound on $R_B(d)$ given by
%If we loose the requirement of classical channels $W$ to classical-quantum channels $\cW : X \ra \cQ$,
%paralleling the conventional discussion of~\cite{CT91, WZ76}, we can have a ``single-letterized''
%expression  
$$
R^*_B(d) = \min_{\cW : X \ra C} \min_{\cD : C B \ra \hat{X}} (I(X; C) - I(C; B))~,
$$ 
where $C$ is now a quantum system (replacing $Y$) and $\cW : X \ra C$ is a classical-quantum channel
(replacing $W$).
Unfortunately, $R^*_B(d)$ appears not to be achievable without entanglement. 
For instance, in the $d = 0$ and $B = \rm{null}$ case, simulating  the channel $X \ra C$ with a rate of 
$I(X;C)$ bits of communication generally requires $H(C)$ ebits ~\cite{BHLSW03}.
Since entanglement cannot be ``derandomized'' like common randomness, a coding theorem
paralleling that of Theorem 4.2 seems unlikely.
 
\section{Bounds on quantum state redistribution}

Our channel simulation with side information result,  Theorem \ref{thm1}, is only partly quantum.
To formulate a fully quantum version of it,
we (i) replace the classical channel $W$ by a quantum feedback channel \cite{Devetak05}
$U^{A \rightarrow \hat{B}\hat{A}}$, which is an isometry from Alice's system $A$
to the system $\hat{B}\hat{A}$ shared by Alice and Bob;
(ii) replace the classical-quantum state $\rho^{XB}$ by a pure state 
$\ket{\varphi}^{RAB}$ shared among the reference system, Alice and Bob.
Sending the $A$ part of $\ket{\varphi}^{RAB}$ through the channel $U$ results in
the state 
$$
\ket{\psi}^{R\hat{A} \hat{B} B} = U \ket{\varphi}^{RAB},
$$
where $\hat{A}$ is held by Alice and $\hat{B}B$ is held by Bob.
Because $U$ is an isometry, the state $\ket{\varphi}^{RAB}$ is equivalent to 
$\ket{\psi}^{R\hat{A} \hat{B} B}$ with $\hat{A}\hat{B}$ in Alice's possession.
Thus simulating the channel $U$ on $\ket{\varphi}^{RAB}$ is equivalent to
\emph{quantum state redistribution}: Alice transferring the $\hat{B}$ 
part of her system $\hat{A}\hat{B}$ to Bob.
We can now ask about the trade-off between qubit channels $[q \ra q]$ 
and ebits $[q \, q]$ needed to effect quantum state redistribution.
In terms of resource inequalities, we are interested in 
the rate pairs $(Q,E)$ such that 
\be
\begin{split}
\< U_1^{S \rightarrow AB} : \rho^{S} \> & + 
 Q \, [q\ra q] +  E \, [q \, q] \\
&\stackrel{s}{\geq} 
\< U_2^{S \rightarrow A\hat{A}\hat{B}} : \rho^{S} \>.
\label{glava2}
\end{split}
\ee
Here $U_1$ is an isometry such that $\ket{\varphi}^{R A B} = U_1 \ket{\phi}^{RS}$, 
$\ket{\phi}^{RS}$ is a purification of $\rho^S$,  and $U_2 = U \circ U_1$.

We can find two rather trivial inner bounds (i.e. achievable rate pairs) based on previous results.
First let us focus on making use of  Bob's side information $B$.
The feedback channel simulation will be performed naively: 
Alice will implement $U^{A \rightarrow \hat{A}\hat{B}}$ 
locally and then ``merge'' her system $\hat{B}$ with Bob's
system $B$, treating $\hat{A}$ as part of the reference system $R$.
This gives an achievable rate pair of 
$(Q_1,E_1) = (\frac{1}{2} I(\hat{B};R\hat{A}), -\frac{1}{2} I(B;\hat{B}))$
 by the fully quantum Slepian-Wolf (FQSW)
protocol \cite{ADHW06,Devetak05}, a generalization of \cite{HOW05}.
The negative value of 
$E$ means that entanglement is \emph{generated}, rather than consumed.

Now let us ignore the side information and focus on performing
the channel simulation non-trivially. This is the domain of
the fully quantum reverse Shannon  (FQRS) theorem \cite{ADHW06,Devetak05,DHLS06}.
Treating $B$ as part of the reference system $R$, the FQRS theorem implies
an  achievable rate pair of 
$
(Q_2,E_2) = (\frac{1}{2} I(\hat{B};RB), \frac{1}{2} I(\hat{B};\hat{A})).
$

%Ideally, one would like to perform a non-trivial channel simulation while
%making use of Bob's side information and we conjecture the following theorem holds.    

\vspace{2mm}

An outer bound is given by the following proposition.

\begin{proposition}
\label{prop:yard}
The region in the $(Q, E)$ plane defined by
$$
Q \geq \frac{1}{2} I(\hat{B};R|\hat{A}), \quad  Q + E \geq H(\hat{B}|B) 
%\label{eq:QE}
$$
contains the achievable rate region for quantum state redistribution.
\end{proposition}

\begin{proof}
Assume that Alice holds $\hat{A} \hat{B}$ and Bob holds $B$. 
Alice wants to transfer her system $\hat{A} \hat{B}$ to Bob.
By the converse to  FQSW (cf. \cite{ADHW06}), transferring $\hat{A}\hat{B}$ requires 
a rate pair $(Q'',E'')$ such that
\be
Q'' \geq \frac{1}{2} I(\hat{B}\hat{A};R), \quad Q'' + E'' \geq  H(\hat{A}\hat{B}|B) .\label{oneshot}
\ee
Now let us perform the redistribution successively: first transfer $\hat{B}$ and 
then $\hat{A}$. 
Let the cost of transferring $\hat{B}$ be $(Q, E)$, which we are trying to bound. 
By FQSW, the cost of transferring the remaining $\hat{A}$ once Bob has $\hat{B}$ can be 
achieved with the rate pair $(Q', E')$ such that
$$
Q' = \frac{1}{2} I(\hat{A};R), \quad Q' + E' = H(\hat{A} | B\hat{B}) .
$$
If $Q < \frac{1}{2} I(\hat{B};R|\hat{A})$,  then
$Q + Q' < \frac{1}{2} I(\hat{B}\hat{A};R)$, which contradicts ($\ref{oneshot}$).
Hence $Q \geq \frac{1}{2} I(\hat{B};R|\hat{A})$ must hold. Similarly, we can prove that
$Q + E \geq H(\hat{B}|B)$.
\end{proof}

\vspace{2mm}

The bound $Q + E \geq H(\hat{B}|B)$ is the analogue of 
the classical bound $R + C \geq H(Y|B)$
from Theorem \ref{thm1}. When $\hat{A} =  {\rm null}$ (simulated channel is the identity) 
the outer bound is achieved by the FQSW-based scheme and when $B = {\rm null}$ (no side information) 
it is achieved by the FQRS-based scheme.  

\section{Discussion}

We have shown here a generalization of both the classical reverse Shannon theorem, 
and the classical-quantum Slepian-Wolf (CQSW) problem. Our main result
is a new resource inequality (\ref{glava}) for quantum Shannon theory.
Unfortunately we were not able to obtain it by naively combining
the reverse Shannon and CQSW resource inequalities via the resource calculus of \cite{DHW05}.
Instead we proved it from first principles. An alternative proof involves 
modifying the reverse Shannon protocol to 
``piggy-back'' independent classical information at a rate of $I(Y;B)$ (cf. \cite{DS03}). 
In \cite{DHW05} certain general principles were proved, such as the ``coherification rules'' 
which gave conditions for when classical communication could be replaced
by coherent communication. It would be desirable to formulate a ``piggy-backing rule''
in a similar fashion.
%This would greatly facilitate the study of trade-offs in quantum information theory,
%which always seem to rely on some form of the reverse Shannon theorem.

An immediate corollary of our result is channel simulation with \emph{classical} side 
information. Remarkably, this purely classical protocol is the basic primitive which 
generates virtually all known classical multi-terminal source coding theorems, not just
the Wyner-Ziv result \cite{LDB06}. 

Regarding the state redistribution problem of Section 5, our results have
inspired Devetak and Yard \cite{DY06} to prove the tightness of the outer bound given by
Proposition \ref{prop:yard}, thus providing the first operational interpretation of
quantum conditional mutual information.
 
\medskip

{\bf Acknowledgement} \,\, This work was supported in part by the NSF grants CCF-0524811 and
CCF-0545845 (CAREER).

%%%%%%%%%%%%%%%%%%%%%%%%%%%%%%%%%%%%%%%%%%%%%%%%%%%%%%%%%%%%%%%%%%%%%%%%%%%%%%%%%%%%%%%%%%%%%%%%%%%%%%%%%%%%%%%%%%%%%%%%
\appendix 

\section{Typicality and conditional typicality}
We follow the standard presentation of \cite{CK81}.
The probability distribution $P_{x^n}$ defined by $P_{x^n}(x) =  \frac{N(x|x^n)}{n}$ is called the \emph{empirical 
distribution} or \emph{type} of the sequence  $x^n$, where $N(x|x^n)$ counts the number of occurrences of $x$ in the word $x^n=x_1x_2...x_n$. A sequence $x^n \in \cX^n$ is called \emph{$\delta$-typical} with respect to a probability distribution $p$ defined on $\cX$ if 
\be
\left| P_{x^n}(x) - p(x) \right| \leq p(x)\delta,  \,\,\, \forall x \in \cX.
\label{eq:typicality}
\ee
The latter condition may be rewritten as
$$
  P_{x^n} \in [p(1 - \delta), p(1 + \delta)].
$$
The set $\cT^{n}_{p, \delta} \subseteq \cX^n$ consisting of all $\delta$-typical sequences is called the $\delta$-typical set. When the distribution $p$ is associated with some random variable $X$, we may use the notation $\cT^n_{X,\,\delta}$. Observe that \eq{typicality} implies 
$$
\| p - P_{x^n} \|_1 \leq  \delta.
$$

The properties of typical sets are given by the following theorem :
\begin{theorem}\label{a1}
For all $\epsilon>0$, $\delta>0$ and sufficiently large $n$, 
\begin{enumerate}
\item $2^{-n [H(p) + c \delta]} \leq p^n({x^n}) \leq 2^{-n [H(p) - c \delta]}$ 
for $x^n \in \cT^{n}_{p, \delta}$,
\item $p^n( \cT^n_{p, \delta}) =  \Pr \{X^n \in \cT^n_{p, \delta} \} \geq 1 - \epsilon$ 
\item $(1 - \epsilon) 2^{n [H(p) - c \delta]} \leq |\cT^n_{p, \delta} |
\leq  2^{n [H(p) + c \delta]}$.
\end{enumerate}
for some constant $c$ depending only on $p$. Above, the distribution $p^n$ is
naturally defined on $\cX^n$ by $p^n({x^n}) = p(x_1) \dots p({x_n})$.
\end{theorem}

\vspace{2mm}

Given a pair of sequences $(x^n,y^n) \in \cX^n \times \cY^n$, the probability distribution $P_{y^n|x^n}$ defined by 
$$
P_{y^n|x^n}(y|x) =  \frac{N(xy|x^ny^n)}{N(x|x^n)}
= \frac{P_{x^ny^n}(x,y)}{P_{x^n}(x)}
$$
is called the \emph{conditional empirical distribution} or \emph{conditional type} of the sequence 
$y^n$ relative to the sequence $x^n$.
A sequence  $y^n = y_1 \dots y_n \in \cY^n$ is called \emph{$\delta$-conditionally typical} 
with respect to the conditional probability distribution $Q$ and a sequence $x^n = x_1 \dots x_n \in \cX^n$ if
$$
  P_{y^n|x^n}(y|x) \in [(1 - \delta)Q(y|x),(1 + \delta)Q(y|x)],   \,\,\, \forall x \in \cX, \forall y \in \cY.
$$
The set of such sequences is
denoted by  $\cT^n_{Q, \delta}(x^n) \subseteq \cY^n$. When $Q$ is associated 
with some conditional random variable $Y|X$, we may use the notation $\cT^n_{Y|X,\,\delta}(x^n)$. 
Define $q(y) = \sum_x Q(y|x) p(x)$.

\begin{theorem}
\label{a2}
For all $\epsilon>0$, 
$\delta > 0$, $\delta'>0$,
and sufficiently large $n$, for all $x^n \in \cT^n_{p, \delta'}$,
\begin{enumerate}
\item  $2^{-n [H(Y|X) + c \delta + c' \delta']} \leq Q^n(y^n|x^n) 
\leq 2^{-n [H(Y|X) - c \delta - c' \delta' ]}$ 
for $y^n \in \cT^{n}_{Q, \delta} (x^n)$.
\item  $Q^n( \cT^n_{Q, \delta}(x^n)| x^n) =  
\Pr \{Y^n \in \cT^n_{Q, \delta}(x^n) | X^n =  x^n\} \geq 1 - \epsilon$ 
\item  $(1 - \epsilon)  2^{n [H(Y|X) - c \delta - c' \delta' ]}
 \leq |\cT^{n}_{Q, \delta} (x^n) |
\leq  2^{n [H(Y|X) + c \delta + c' \delta' ]} $.
\item If $y^n \in \cT^{n}_{Q, \delta}(x^n)$, then $(x^n, y^n) \in \cT^{n}_{pQ, (\delta + \delta' + \delta \delta')}$
, and hence $y^n \in  \cT^{n}_{q, (\delta + \delta' + \delta \delta')}$.
\item  $Q^n( \cT^n_{q,\delta + \delta' + \delta \delta' }| x^n)  \geq 1 - \epsilon$. 
\end{enumerate}
for some constants $c, c'$ depending only on $p$ and $Q$. 
\end{theorem}

\bibliography{ref}
\bibliographystyle{abbrv}

\end{document}